# Computer-aided detection of pulmonary nodules in low-dose CT

P. Delogu & M.E. Fantacci
*Dipartimento di Fisica dell'Università & INFN, Pisa, Italy*

I. Gori
*Bracco Imaging S.p.A., Milano, Italy*

A. Preite Martinez
*Centro Studi e Ricerche Enrico Fermi, Roma, Italy*

A. Retico
*Istituto Nazionale di Fisica Nucleare, Sezione di Pisa, Italy*

ABSTRACT: A computer-aided detection (CAD) system for the identification of pulmonary nodules in low-dose multi-detector helical CT images with 1.25-mm slice thickness is being developed in the framework of the INFN-supported MAGIC-5 Italian project. The basic modules of our lung-CAD system, a dot-enhancement filter for nodule candidate selection and a voxel-based neural classifier for false-positive finding reduction, are described. Preliminary results obtained on the so-far collected database of lung CT scans are discussed.

## 1 INTRODUCTION

Lung cancer is one of the most relevant public health problems. Despite significant research efforts and advances in the understanding of tumor biology, there has been no reduction of the mortality over the last decades. Lung cancer most commonly manifests itself with the formation of non-calcified pulmonary nodules. Helical Computed Tomography (CT) is recognized as the best imaging modality for the detection of pulmonary nodules (Diederich et al. 2001). The amount of data that need to be interpreted in CT examinations can be very large, especially when thin collimation settings are used, thus generating up to about 300 2-dimensional images per scan, corresponding to about 150 MB of data.

In order to support radiologists in the identification of early-stage pathological objects, researchers have recently begun to explore computer-aided detection (CAD) methods in this area.

The First Italian Randomized Controlled Trial aiming at studying the potential impact of screening on a high-risk population using low-dose helical CT has recently started (Italung-CT trial). In this framework and in the framework of the MAGIC-V (INFN CSN-V) project, we are developing a CAD system for pulmonary nodule identification, intended to be used as a supportive tool in screening protocols. The system is based on a dot-enhancement filter and a neural classifier for the reduction of the number of false-positive (FP) findings. This paper describes a comparison between the performances of two different voxel-based neural approaches originally developed for the classification of the selected region of interest (ROI) by the dot-enhancement filter.

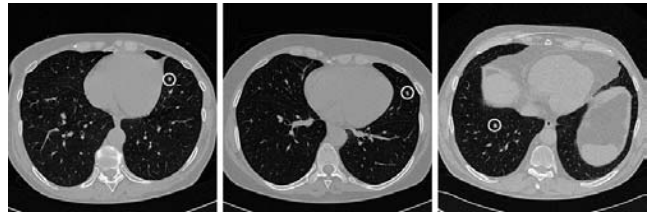

Figure 1. Some examples of pulmonary nodules.

## 2 THE CAD SYSTEM ARCHITECTURE

Pulmonary nodules may be characterized by very low CT values and/or low contrast, they may have CT values similar to those of blood vessels and airway walls to which they could also be strongly connected. Examples of lung internal nodules are shown in Figure 1.

An important and difficult task in the automated nodule detection is the initial selection of the nodule candidates within the lung volume. In the identification process, lung nodules are modeled as spherical objects and a dot-enhancement filter is applied to the 3D matrix of voxel data. The filter attempts to determine the local geometrical characteristics of each voxel, by computing the eigenvalues of the Hessian matrix and evaluating a likelihood function that was specifically built to discriminate between the local morphology of linear, planar and spherical objects, the latest modeled as having 3D Gaussian sections (Li et al. 2003; Delogu et al. 2005). A simple peak-detection algorithm (i.e.

a local maximum detector) is then applied to the filter output to detect the filtered-signal peaks.

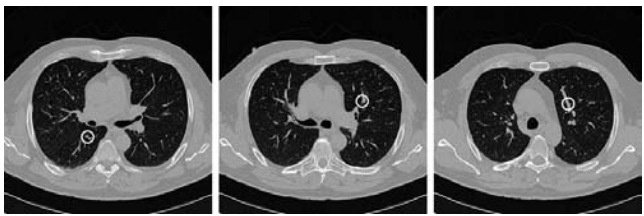

Figure 2. Some examples of false positive findings generated by the dot-enhancement filter.

Since most FP findings are provided by crossings between blood vessels (see Fig. 2), we attempted to reduce the amount of FP/scan by developing a procedure which we called voxel-based neural approach (VBNA). According to that method, each voxel of a region of interest (ROI) is characterized by the grey level intensity values of its neighbors (see Fig. 3). We developed, implemented and compared two different VBNA procedures. In the first, the CT values of the voxels in a 3D neighborhood of each voxel of a ROI are rolled down into vectors of features (147 features) to be analyzed by a neural classifier. In the second procedure (Gori, I. & Mattiuzzi, M. 2005), 6 additional features constituted by the eigenvalues of the gradient and the Hessian matrices are computed for each voxel and encoded to the feature vectors (153 features). A feed-forward neural network is implemented at this stage to assign each voxel either to the nodule or normal tissue target class.

A candidate nodule is then characterized as "CAD nodule" if the number of pixels within its ROI tagged as "nodule" by the neural classifier is above some relative threshold. A free response receiver operating characteristic (FROC) curve for our CAD system can therefore be evaluated at different threshold levels.

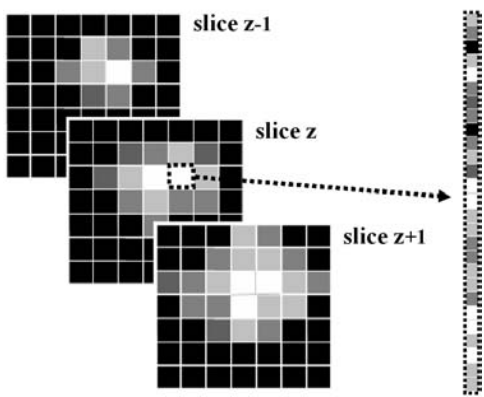

Figure 3. Voxel-based neural approach to false-positive reduction.

## 3 DATA ANALYSIS AND RESULTS

The CAD system was developed and tested on a dataset of low-dose (screening setting: 140 kV, 70÷80 mA) CT scans with reconstructed slice thickness of 1.25 mm. The scans were collected and annotated by experienced radiologists in the framework of the screening trial being conducted in Italy (Italung-CT). The database available for this study consists of 14 scans, containing 24 internal nodules. Each scan is a sequence of about 300 slices stored in the DICOM (Digital Imaging and COmmunications in Medicine) format.

First of all, the lung volume is segmented out of the whole 3D data array by means of a purposely built segmentation algorithm that identifies the internal region of the lung (Antonelli et al. 2005).

The 3D dot-enhancement filter applied to the selected lung regions shows a very high sensitivity. In particular, the lists generated by the peak-detector algorithm for all CT are empirically truncated so to include all annotated nodules. According to this procedure, a 100% sensitivity to internal nodules is obtained at a maximum (average) number of 54 (52.3) FP/scan.

With respect to the VBNA procedure for FP reduction, the dataset was randomly partitioned into train and test sets; the performances of the trained neural networks were evaluated both on the test sets and on the whole dataset.

In the first VBNA approach, 147 features, derived from a 2D region of 7x7 voxels for 3 consecutive slices with the voxel to be classified in the center, constitute each vector of the feature dataset. Two three-layer feed-forward neural networks with 147 input, were trained on two different random partitions of the dataset into train and test sets. The performances achieved in each trial for the correct classification of individual pixels are reported in Table 1, where the sensitivity and the specificity values obtained on the test sets, on the whole datasets and the average values on the two trials are shown.

Table 1. VBNA with 147 features

| test | | train+test | |
|---|---|---|---|
| sens % | spec % | sens % | spec % |
| 71.7 | 82.7 | 81.5 | 87.6 |
| 73.4 | 78.1 | 78.1 | 80.6 |
| | average | 79.8 | 84.0 |

Also in case of the second VBNA approach, when 6 additional features are encoded to each feature vector, two different neural networks were trained, obtaining the performances shown in Table 2.

Table 2. VBNA with 153 features

| test | | train+test | |
|---|---|---|---|
| sens % | spec % | sens % | spec % |
| 75.3 | 78.6 | 85.5 | 83.3 |
| 79.9 | 84.1 | 84.6 | 88.1 |
| | average | 85.1 | 85.7 |

The comparison between the average sensitivity and specificity obtained in the first and second approach proves that the implementation of some features exploiting the morphology of the voxel neighborhood in addition to the textural features directly derived by the sequence of the voxel intensity values can improve the system discriminating capability.

Once the VBNA approach with 153 features has been applied to each ROI, the FROC curves have been evaluated for the two trained neural networks. They are shown in Figure 4. Both curves show the 100% sensitivity at a very low level of false positives (1.5-3.5). The sensitivity value remains very high (85% range) even at less than 1 false positive per scan.

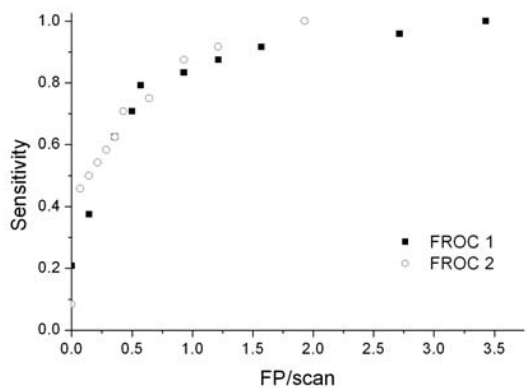

Figure 4. FROC curves obtained on a dataset on 14 CD scans containing 24 internal nodules.

## 4 CONCLUSIONS

In this work we compared two different voxel-based neural approaches (VBNA) to the FP reduction in the framework of the development of a CAD system for the identification of pulmonary internal nodules in low-dose CT scans. The VBNA approach where for each voxel 6 morphological features have been added to the vector of rolled-down voxel neighborhood, has shown the best FP reduction capability.

In conclusion, the dot-enhancement pre-processing algorithm provides a good sensitivity in the identification of nodule candidates and the VBNA with 153 features is an effective approach to the problem of false positives reduction. In particular, the VBNA approach allows to reduce the average rate of 52.3 FP findings per scan generated by the dot-enhancement filter to 1.9 FP/scan for a sensitivity of 100% (24 nodules detected out of 24). If the sensitivity value is decreased to 87.5% (21 nodules detected out of 24), a rate of 0.9 FP/scan is obtained. These preliminary results are promising, albeit a validation against an independent validation database is required.

ACKNOWLEDGMENTS

We thank the *MAGIC-5 Collaboration (INFN, CSN-V)* for contributing to this research. We acknowledge Dr. L. Battolla, Dr. F. Falaschi and Dr. C. Spinelli of the U.O. Radiodiagnostica 2 dell'Azienda Ospedaliera Universitaria Pisana and Prof. D. Caramella and Dr. T. Tarantino of the Divisione di Radiologia Diagnostica e Interventistica del Dipartimento di Oncologia, Trapianti e Nuove Tecnologie in Medicina dell'Università di Pisa for providing the annotated database of CT scans. We are grateful to Dr M. Mattiuzzi from Bracco Imaging S.p.A. for useful discussions.